\documentstyle[12pt]{article}
\author{A.  Jadczyk \\
Institute of Theoretical Physics,
 University of Wroc{\l }aw,  \\
Pl.  Maxa Borna 9,  PL 50 204 Wroc{\l }aw,  Poland}
\title{On Quantum Jumps, Events and Spontaneous Localization Models\\
bulletin board ref.: hep-th/9408021}
\date{ }
\def\hq{{\cal H}_q}
\def\rh{{\hat\rho}}
\def\cs{{\cal S}_{cl}}
\def\be{\begin{equation}}
\def\ee{\end{equation}}
\def\ba{\begin{array}} \def\ea{\end{array}} 
\def\tr{\mbox{Tr}\, }
\def\lqq{\lq\lq}
\def\rqq{\rq\rq}
\begin{document}
\maketitle
\begin{abstract}
We propose a precise meaning to the concepts of \lqq experiment\rq\rq,
\lqq measurement\rqq\,  and \lqq event\rqq\,  in the event-enhanced
formalism of quantum theory. A minimal piecewise deterministic process is
given that can be used for a computer simulation of real time series of
experiments on single quantum objects. As an example a generalized cloud
chamber is described, including multiparticle case. Relation to the GRW
spontaneous localization model is discussed.
\end{abstract}
\section{Events, Measurements and Experiments}
In his paper \lqq The Philosophy of Experiment\rqq\,  E.
Schr\"odinger$^{(1)}$
wrote:
\begin{quotation}
\lqq  The new science (q.m.) arrogates the right to bully our whole
philosophical outlook. It is pretended that refined measurements which
lend themselves to easy discussions by the quantum mechanical formalism
could actually be made. (...) Actual measurements on single individual
systems are never discussed in this fundamental way, because the theory
is not fit for it.(...) We are also supposed to admit that the extent of
what is, or might be, observed coincides exactly with what the quantum
mechanics is pleased to call observable.\rqq
\end{quotation}
It is well known to every experimentalist that more can be, and {\sl is}
being observed than q.m. is pleased to call an observable. What
Schr\"odinger asserted 40 years ago is even more valid today. Actual
measurements are nowadays often being made on single quantum objects.
They give us finite time series of {\sl events}. The importance of this
concept of an \lqq  event\lqq , and the intrinsic incapability of quantum
theory to
deal with it, have been stressed by several authors. H.P. Stapp$^{(2,3,4)}$
emphasized the role of \lqq events\rqq\,  in the \lqq world process\rqq. G.
F. Chew used
Stapp's$^{(5)}$ ideas on soft--photon creation--annihilation processes
and proposed$^{(6)}$ the term \lqq explicate order\rqq, complementing Bohm's
\lqq implicate\rqq\,  quantum order, to denote the world process of \lqq
gentle\rqq\,
creation--annihilation events. A. Shimony$^{(7)}$ accentuated \lqq  that
quantum mechanics does not have, in any obvious way, the conceptual tools for
explaining how potentialities are actualized\rqq. He thought a solution in
refining the concept of event and conjectured that \lqq the actualization of
potentiality must not be conceived as a limiting case of probability -
as probability 1 or 0. Instead, actuality and potentiality are radically
different modalities of reality\rqq. R. Haag$^{(8)}$ emphasized that \lqq
an event in quantum physics is discrete and irreversible\rqq\, and that \lqq
we must assume that the arrow of time is encoded in the fundamental laws
...\rqq.
He also suggested$^{(9)}$ that \lqq  transformation of possibilities into
facts must be an essential ingredient which must be included in the
fundamental formulation of the theory\rqq.\footnote{A.O. Barut$^{(10)}$
seeked also a theory of events, but in terms of \lqq wave lumps\rqq\,
rather
than in terms of discrete collision--like or detection--like discrete
occurences}\\
J.S. Bell$^{(11,11a)}$ reprimanded the misleading use of the term
\lqq measurement\rqq\,  in quantum theory. He opted for banning this word
from
our
quantum vocabulary, together with other vague terms such as
\lqq macroscopic\rqq, \lqq microscopic\rqq, \lqq observable\rqq\,  and
several others. He suggested to replace the term \lqq measurement\rqq\,  by
that of \lqq experiment\rqq, and
also not to speak of \lqq observables\rqq\,  (the things that seem to call
for an
\lqq observer\rqq) but to introduce instead the concept of \lqq
beables\rqq  - the
things that objectively \lqq happen--to--be (or
not--to--be)\rqq.\footnote{Calling observables \lqq observables\rqq\,  can
be,
however, justified in the event-enhanced formalism that we are outlining
here.} But there is no place for \lqq events\rqq\,  or for \lqq
beables\rqq\, in ordinary
quantum theory. That is so because each \lqq event\rqq\,  must have three
characteristic features:
\begin{itemize}
\item it is classical,
\item it is discrete,
\item it is irreversible.
\end{itemize}
If just one of these three features is relaxed, then what we have is not
yet an \lqq event\rqq.\\
It must be {\sl classical}, because it must obey to the classical
\lqq yes-no\rqq\,  logic; it must never be in a \lqq superposition\rqq\,  of
it being happened and being unhappened. Otherwise it would not be an event.\\
It must be {\sl discrete}. It must happen wholly. An event that \lqq
approximately\rqq\, happened is not an event at all.\\
It must be {\sl irreversible}, because it can not be made \lqq undone\rqq.
This feature distinguishes real evnts from the \lqq virtual\rqq\,  ones. Once
something happened -- it happened at a certain time instant. It must
have left a trace. Even if this trace can be erased, the very act of
erasing will change the future -- not the past. Something else may
happen later, but it will be already a different event. We believe
that {\sl the events, and nothing but events, are pushing forward the
arrow of time}.

Once these three characteristics of an event are accepted (and they are
evident for an experimentalist), it becomes clear what is necessary if
we want to enhance the standard formalism of quantum theory so as to
include events into it.\\

First -- we must allow the formalism to include {\sl classical} quantities.
We believe that
is better when this is done openly rather than via a back door.\\
Second -- we must allow for an {\sl irreversible} coupling between quantum
and classical degrees of freedom. It must be stressed that the minimal
irreversibility that we are talking about here {\sl is not} the result of
\lqq noise\rqq\, \lqq chaos\rqq\, or \lqq environment\rqq, but is forced
upon by a universal law: the one stating that: {\sl information must be
paid with dissipation}.\\ But that is not enough. We need to make
another important step: we must learn how to describe {\sl finite time series
of events} that are registered in experiments. It is out of such time
series that expectation values can be computed. But the experimentalist
is interested not only (or not at all) in expectation values. (He may be,
for instance, interested in time correlations for a finite sequence of
neutron detection events$^{(12,13)}$.) Moreover, as human beings, we are
interested perhaps only in one such time series of events at all -- as we
cannot enter twice into the same stream of time. Not only we want to be
able to compute statistical characteristics of ensembles. We also want
to be able to simulate on digital computers experimental finite time
series for individual systems. We want to be able to simulate the events
that form up \lqq mini-universes\rqq\, governed by the same laws as that
great one that we live in. We want to know how to account for all
regularities that are apparently seen in the acquired data. We also seek
guiding principles that will tell us what to do if we want to see still
more regularities in the workings of Nature that are being unveiled to
us.

Can quantum theory be tailored so as to suit these demands? Or, perhaps,
it suits them already? It is not our aim to describe all the efforts
that has been taken by quantum physicists in last 70 years in this
direction. Scanning through the recent physical literature will give the
reader an idea of what was done, what progress has been achieved. In
this paper we just want to describe what seems to us to be {\sl the
minimal enhancement} of quantum theory that suits the demands of human
experience and technology. It has been developed in a series of papers
by Ph.Blanchard and the present author$^{(14-21)}$. It is one of the aims of
the present paper to express, in a condensed form, the \lqq philosophical
backbone \rqq\, that can be felt through the mathematical skeen of
the several model discussed in Refs (14-21).
{}From the structural
and from mathematical point of view, the three most essential features
of this enhanced formalism are:
\begin{itemize}
\item tensoring of non--commutative quantum algebra with a commutative
algebra of functions,
\item replacing Schr\"odinger's unitary dynamics of pure states with a
suitable completely positive time evolution of families of density
matrices,
\item interpreting the continuous time evolution of statistical states
of the total (i.e. quantum+classical) system in terms of a piecewise --
deterministic random process on pure states. The process consists of
pairs (quantum jump,classical event) interspersed by random periods of
Schr\"odinger-type, in general non--linear, continuous evolution.
\end{itemize}
In an informal way that extension can be understood as follows. When
describing any real (i.e. actual, not some imagined one) quantum system
$Q$, we always have to decide at which point our quantum description
ends. We cannot include \lqq everything\rqq\,  into $Q$. That for the simple
reason that we do not know what \lqq everything\rqq\,  is (tomorrow we will
perhaps know more than we know today). Then, where $Q$ ends, the new
territory begins. We may ignore this new territory completely and
restrict ourselves to $Q$. Or, we can try to take it also into account.
This new territory may include \lqq environment\rqq, may include
\lqq events\rqq, but
it may include also \lqq us\rqq\,  - who are creating theories, making
predictions, and who are checking those theories by doing experiments.
Taking into account environment is standard and does not need a
separate discussion. On the other hand discussing human
minds and acquisition of human knowledge would take us too far, and to
a shaky ground.
The present version of the formalism, even after \lqq enhancement\rqq\,  as
sketched in the present paper, does not seem to be powerful enough to
carry out this last step. So, we restrict ourselves to the
decription of events. The logic that we are necessarily
using
when
checking our reasoning against errors, and when describing experimental
events -- is the classical logic of Aristotle and
Boole. This classical description can, and always partially will, extend
downwards: from minds to brains, to senses, to communication channels,
and still down to measuring devices. What results from such an extension
constitutes a classical system $C$. Of course we may and we should use,
whenever it helps, the knowldge that we have about
microscopic structure of instruments, light quanta and brains, and we
may appropriately enlarge $Q$. But in each case there will be a limit.
Going beyond that limit will be useless and often impractical if our
aim is to understand a given physical phenomenon rather than  to  construct
an abstract \lqq theory of  everything\rqq\,  that computes and explains
nothing.
Anyhow, whatever we choose to include into $Q$, at some point we will
have to decide what constitutes for us the \lqq events\rqq\,  that we want
to account for.  It is the believe of the present author
that {\sl an event can be only defined as \lqq a change of a state of
$C$\rqq.} As an example we may think of $Q$ as being a single particle, for
instance an electron, and of $C$ as a particle detector that can be in
one of its two states: \lqq off\rqq\,  (particle undetected) or \lqq
on\rqq\, (particle detected). Then an \lqq experiment\rqq, or a \lqq
measurement\rqq, is nothing but an
appropriate coupling of $Q$ and $C$ together, observing $C$, and
learning from it about $Q$. Or, going to the extreme end, we can think
of $Q$ including \lqq everything\rqq\,  but \lqq minds of the
observers\rqq\, $^{(22)}$.
As noticed by Heisenberg$^{(23)}$, denying necessity of such a cut, and
attempting to make all the universe into a single quantum system,
leads to a situation when \lqq physics has vanished, and only mathematical
scheme remains\rqq. We would go even further and add:
but even not that, because this mathematical scheme is
then also a part of the quantum universe, and so it is only one of the
unimaginable infinity of other potentially possible schemes. \\
On the practical side, we believe that our approach, after suitable
translation, can be acceptable even by the quantum purists who probably
would like to deny the fact that there are any
{\sl classical} events. First of all let us point it out that they are using,
perhaps without noticing it, classical events  in the form of
{\sl position of the state vestor
in the space of all possible state vectors}. An attempt of
\lqq quantizing\rqq\,
even this position would lead to an absurd infinite sequence of
quantizations,
and thus to a theory that looses completely contact with reality.
So, even if \lqq wave function is observable, after all\rqq, then {\sl it
is a classical observable}.
Second, quantum purists may consider our $C$ as a copy of their
\lqq pointer basis\rqq. Then perhaps
they will notice that our dynamics always respects this basis. It is
for this reason that we may call it consistently classical.
They will also notice that in  our approach,  it is only in special
cases that the effective evolution of reduced quantum density
matrix separates -- which makes our approach more general. Finally,
we do not mind calling our approach \lqq phenomenological\rqq\,  or \lqq
effective\rqq.
In fact we believe that any theory that has anything to do with
the particular reality that is given to us, must be both phenomenological
and effective to a necessary degree.

Let us discuss now the two terms: \lqq measurement\rqq\,  and
\lqq experiment\rqq. The
term \lqq experiment\rqq\,  seems to be less dangerous than that of
\lqq measurement\rqq\,
-- the last one being discredited owing to its over--fuzziness and
ar\-bi\-tra\-ri\-ness.$^{(10,11)}$ Moreover, from any actual coupling
between a quantum system and measuring and recording devices, from any
actual experiment, we can usually learn many different things by
postselection of data and by numerical analysis of the data
afterwards.$^{(12,13)}$ As it was stressed by E. Schr\"odinger in the
opening quotation, an actual experiment can rarely -- if ever -- be
considered as a measurement of a quantum mechanical \lqq observable\rqq.
According to already quoted reprimanding papers by J. Bell, the term
\lqq observable\rqq\,  should be banned -- together with that of
\lqq measurement\rqq\,  and
a couple of others. We feel that rather than ban, it is better, whenever
possible, to define the terms precisely. We propose the following
definitions:
\newtheorem{definition}{Definition}
\begin{definition}
An {\bf experiment} is a CP coupling between a quantum and a classical
system. One observes then the classical system and attempts to learn
from it about characteristics of state and of dynamics of the quantum
system.
\end{definition}

\begin{definition}
A {\bf measurement} is an experiment that is used for a particular
purpose: for determining values, or statistical distribution of values,
of given physical quantities.
\end{definition}

By {\sl CP coupling}\, we mean a two--parameter family $\phi_{t,s}$,
$t\leq s$, of completely positive maps on the algebra of $C+Q$,
satisfying the composition law
\be
\phi_{t,s}\phi_{s,r}=\phi_{t,r} ,
\label{eq:comp}
\ee
and
\be
\phi(t,t)=Id
\label{eq:tt}
\ee
If $\phi_t$ acts on operators, then speaking of a \lqq measurement\rqq, we
will usually require
\be
\phi_{t,s}(I)=I.\label{eq:id}
\ee
\noindent
{\bf Remark 1.} We can also think of experiments that include state
preparation (and thus selection or filtering) procedures. In that
case the condition (\ref{eq:id}) should be relaxed. In the present paper
we will not discuss state preparation parts of experiments, so we
will always assume Eq. (\ref{eq:id}) to hold.\medskip\\
\noindent
{\bf Remark 2.} When speaking of experiment and of observing the
classical system, we may have in mind observations on a single system
that last in time, or we can allow for repeated observations on
similarly prepared systems. The standard quantum mechanical concept of
measurement, whose analysis goes back to J. von Neumann, deals
exclusively with the second case, while our description allows for
experiments on single systems. That such an extension is necessary is
nowadays clear, as advances in technology make such experiments on
quantum systems more and more frequent. Moreover, we are finding pure
quantum effects in macroscopic systems, and there we are making
prolonged experiments on single objects (e.g. SQUID-s). Finally, in some
experiments, as for example for a SQUID-tank system$^{(24)}$, we may be
changing the quantum dynamics depending on the actual state of our
measuring devices. The event--enhanced formalism easily accomodates
such experiments.$^{(18)}$\medskip\\
\noindent
{\bf Remark 3.} One more important remark is due here. In the definition
above we took it for granted that the classical system $C$ can be
\lqq observed\rqq, and that we do not have to define what does {\sl that}
mean.
The point is that states of a quantum system always change as a result
of an observations. It is for this reason that a \lqq quantum measurement
theory\rqq\,  is necessary. On the other hand, given two classical systems
$C_1$ and $C_2$, we can couple $C_1$ and $C_2$ together in such a way
that the information about the actual state of $C_1$ is translated into
the actual state of $C_2$, and that the change of the actual state of
$C_1$ caused by this coupling is arbitrarily small. In fact, for a
complete consistency, we need something more: we must prove that if
$C_1$ is coupled to $Q$, and $C_2$ is coupled to $C_1$, then this latter
coupling can be such that it will not affect $Q$ otherwise than via
$C_1$. Such a proof may be difficult or impossible if one wants to think
of $C$ as of, for instance, a classical electromagnetic or gravitational
field.

\section{The Minimal Extension of Quantum Theory that Accounts for
\lqq Events\rqq}
The algebraic framework that is needed for a mathematical formulation of
such an extension of quantum theory is described in Refs. (14,19). Here
we will describe it briefly and in plain terms, without aiming at a
mathematical rigour. A formulation using more general, algebraic,
language is also possible, but no significant physical insight would be
gained concerning the problems at hand.

Suppose that we want to describe a quantum system $Q$ coupled to a
classical system $C$. Let $\hq$ be the Hilbert space which is used to
describe $Q$. Pure states of $Q$ are described by unit vectors
$\psi\in\hq$ (modulo phase). Statistical states are described by density
matrices $\rh$, $\rh\geq 0$, $\tr (\rh )=1$. Observables are described
by operators $A\in\hq$, and their expectation values in states are given
by $<A>_\psi =(\psi,A\psi)$, $<A>_\rh =\tr (A\rh )$.\footnote{We should
stress here that the traditional term \lqq expectation value\rqq\,  can be
justified within event-enhanced formalism. It is indeed an expectation
value computed from a sequence of events that result from a special
coupling of $Q$ to appropriate $C$.}

Classical system $C$ is described by a space $\cs$, whose points are
pure states of $C$. For a dynamical system, $\cs$ will usually be the
phase space of $C$. But, if the dynamics of $C$ itself is trivial (as it
will be, for instance, if $C$ is discrete), then $\cs$ can be the
configuration space of $C$ as well. In general $\cs$ should be equipped
with some measurable structure. To make our discussion as simple as
possible, and not to dissipate our attention on mathematical subtleties,
let us assume here that $\cs$ is a finite set, parameterized by a
parameter $\iota$. Thus we will skip free dynamics of $C$.
\footnote{Taking it into account is not a problem. In the SQUID--tank
model$^{(18)}$ $\cs$ is a two-dimensional phase space with dumped
oscillator dynamics.} Statistical states of $C$ are probabilistic
measures on $\cs$ or, in our case, sequences of non-negative numbers
$\mu_\iota$, with $\sum_\iota\mu_\iota=1$. Observables of $C$ are
functions on $\cs$.

We consider now the total system $Q+S$. Its pure states are pairs
$(\psi,\iota)$. Its statistical states are families $\rho_\iota$, such
that, for each $\iota$, $\rho_\iota$ is a positive operator in $\hq$,
and that $\sum_\iota \tr (\rho _\iota)=1$. A statistical state
$\{\rho_\iota\}$ of $C+Q$ determines, by partial tracing, the effective
states ${\rh}$ of $Q$ and $\mu$ of $C$: $\rh=\sum_\iota\rho_i$,
$\mu_\iota=\tr(\rho_\iota)$. Observables of $Q+C$ are families
$A_\iota$. Expectations are given by $<A>_\rho = \sum_\iota \tr
(A_\iota\rho_\iota)$. The allowed couplings of $Q$ and $C$ are described
by the Liouville equation with a (possibly time--dependent)
Lindblad--type generator$^{(25,26,27)}$

We will not need the most general form of such a coupling here, so we
describe only a particular, simple case. It will correspond to the
following {\sl informal} description of the coupling:
\begin{itemize}
\item the coupling does not explicitly depend on time (it is easy to
relax this condition)
\item there is a certain number of \lqq quantum properties\rqq\,  $F_\alpha$,
$\alpha=1,2,\ldots $ of $Q$ that we want to discriminate between (to
\lqq measure\rqq),
\item if the classical system is in a state $\iota$ then the quantum
system evolves according to a Hamiltonian quantum dynamics described by
some quantum Hamiltonian $H_\iota$,
\item for each $\alpha$ there is a certain transformation of $\cs$ with
the following meaning: if the quantum system \lqq has\rqq\,  a property
$F_\alpha$
while the classical system is in a state $\iota$, then $C$ switches from
$\iota$ to a new state, which we denote $\alpha (\iota)$.
\end{itemize}

For simplicity we will assume that the maps
$\alpha : \iota \mapsto \alpha(\iota)$ are
one-to-one. Even more, for the present purpose we will assume that each
of them is an involution, that is that $\alpha(\alpha(\iota))=\iota$.
That assumtions corresponds to the idea that $C$ consists of two-state
subsystems, and that each $\alpha$ is flipping states in some of these
subsystems. Or, in other words, that the classical (pure) states are
described by strings of zero-one bits,
and that every transformation $\alpha$
flips the bits in some (depending on $\alpha$) substring. Thus our
coupling -- although (because of no-go theorems)
 necessarily irreversible -- is only as much
irreversible as demanded by its very nature. The reversible maps
$\alpha$ resemble thus the idea of conservative logic$^{(28)}$.

We have not yet described what we mean by the term \lqq property\rqq\,  of a
quantum system. The simplest example of a property is an orthogonal
projection $e\in\hq$. A {\sl fuzzy} property is a positive operator $a$,
$0\leq a \leq I$. It is convenient to allow for properties to have
different \lqq intensivities\rqq, therefore we relax the assumption
$a\leq I$,
and so
$a$ can be just a positive operator. Thus, what we call a \lqq property\rqq,
corresponds to what is also known, especially in the Ludwig's school,
under the name \lqq effect\rqq. In application $F_\alpha$-s will either
come
from some spectral measure, or from some smoothed-out spectral measures
(so called POV measure) or, as will be the case in the next section,
from smoothed integrals of number density operator over space regions.
Finally, in order to allow for still  wider applications of the formalism,
we need not even to assume that the operators $F_\alpha$ are Hermitian
(although in most applications they will be positive) - then they may
be called \lqq operations\rqq. \footnote{Usually, when modelling a
measurement, the $F_\alpha$ are positive, but there are experiments that
are not measurements (as for instance the experiment performed by
Nature that we are participating in). Then $F_\alpha$ may be creation
or annihilation operators.}

With this heuristic description of the coupling in mind, we will write
now the form of the Liouville equation that corresponds to it. A
mathematical justification can be found in Ref. (15,19); see also
references therein.
\be
{{d\rho_\iota}\over {dt}}=-i\left[H_\iota,\rho_\iota\right]+\sum_\alpha
F_\alpha\rho_{\alpha(\iota)}F_\alpha^{\star}-{1\over
2}\{\Lambda,\rho_\iota\},
\label{eq:rd}
\ee
where we have denoted
\be
\Lambda\doteq \sum_\alpha F_\alpha^{\star} F_\alpha\ .\ee
We can always switch the time evolution between states and observables
by using the duality equation: $<A(t)>_\rho=<A>_{\rho(t)}$. Then for
observables we have almost the same equation as above,
except for the sign in front of the commutator and the order of $F$ and
$F^\star$ in the second term:
\be
{{dA_\iota}\over {dt}}=i\left[H_\iota,A_\iota\right]+\sum_\alpha
F_\alpha^{\star} A_{\alpha(\iota)}F_\alpha-{1\over
2}\{\Lambda ,A_\iota\}.\label{eq:ad}
\ee

We notice now that in a special case, {\sl if the quantum Hamiltonian does
not depend on
the state of the classical system}, i.e. if $H_\iota\equiv H$ for
each $\iota$, then Eq. (\ref{eq:rd}) can be summed up over
$\iota$. This is so because we have assumed that each of the
transformations $\alpha$ is one--to--one and onto (i.e. it is a
permutation in $\cs$). Therefore, for each $\alpha$,
$\sum_\iota\rho_{\alpha(\iota)}=\sum_\iota \rho_\iota=\rh$. It follows
that, in this special case, the time evolution for the effective quantum
states separates and we have:
\be {d{\rh}\over {dt}}=\sum_\alpha F_\alpha\rh
F_\alpha^{\star}-{1\over2}\{\Lambda,\rh\}.
\label{eq:rqd}
\ee
It should be stressed that this separating property of the Liouville
equation that governs time evolution of the total system $C+Q$, need not
to hold for more general models. If the quantum Hamiltonian depends on
$\iota$, or if there is a non-trivial dynamics in the classical system,
then summing over $\iota$-s as above is not possible. That is what
happens e.g. in the SQUID-tank model.\\
We also notice that, even in those cases when the effective quantum
dynamics separates,  Eq.(\ref{eq:rqd}) contains less information than
Eq.(\ref{eq:rd}). Indeed, information about the transformations
$\alpha:\iota\mapsto\alpha(\iota)$, that is information about events, is
lost. Therefore there can be many couplings of $Q$ to other systems
(classical or quantum) that determine effectively the same evolution of
statistical quantum states. Some of these couplings may be totally
useless, for instance if $\alpha(\iota)=\iota$ for all $\alpha$ and
$\iota$; or if the $\alpha$-s are simply mixing $\cs$. In that case
nothing useful happens to the classical system that could be used for
learning about the quantum one. Entropy of the quantum system is growing
with no useful gain. There are either no events and no observations at
all, or the events are chaotic and information is being lost. We have,
in such a case, a useless dissipation in the quantum system (unless our
aim is to build a quantum driven random number generator, which is also
of intertest). On the
contrary to this, {\sl the aim of any well posed measurement is to get a
maximum of information and to pay for it with a minimum of
dissipation.}\\
Because dissipation of quantum states is already fixed in our simplified
model by Eq. (\ref{eq:rqd}) (independently of the choice of involutive
transformations $\alpha$), it is clear that $\cs$ and the maps
associated to $\alpha$-s should be chosen in an optimal way. We will see
a particular example in the next section, where we will formulate the
multi-particle cloud chamber model.\footnote{In the following we will
always assume that each $\alpha$ acts in a non--trivial way on $\cs$.
Without this assumption the following discussion of the stochastic
process would have to be more subtle.}

Till now we were discussing time evolution of statistical states. But,
as it was already pointed out above, in many experiments we are
interested in time series of events. In fact, often these are the only
results that are provided. Sometimes (but not always) the experiment
can be repeated
another time in another place. But even then, this is already another
experiment. Fortunately it so happens that the information contained in
the Liouville generator on the rhs of Eq. (\ref{eq:rd}) is sufficient
for reconstruction of the minimal, piecewise deterministic random
process that describes individual histories and which allows for a
computer simulation of actual experiments. Although a fully satisfactory
mathematical justification of this reconstruction is yet to be given, a
general idea and examples have been described in Refs. (16-21). We will
not repeat the derivation once more. Instead, in the following
subsection, we will describe explicitly the piecewise deterministic
random process (PDP) that corresponds to the Eq. (\ref{eq:rd}), that is
which leads to this equation after averaging over individual
histories.\medskip\\
\noindent{\bf Remark 4}
An easy introduction to the theory of PDP  that does not require
the theory of stochastic differential equations can be found in Refs.
$(29,30)$. Random processes that lead to equations similar to Eq.
(\ref{eq:rqd}) were considered by other authors - cf. Refs. $(31-34)$,
but they were all concerned only with the the
quantum part of evolution. They were not concerned with \lqq events\rqq\,
and with information that can be obtained from time series of events. In
Pearle$^{(35)}$, Ghirardi and Pearle$^{(36)}$ (see also Gisin$^{(37)}$
and references therein) stochastic differential
equations were used that lead to diffusion processes in Hilbert space.
Although reproducing the master equation for the quantum subsystem,
these diffusion processes fail to be equivalent to PDP processes
when applied to $Q+C$. We consider this as an
unsatisfactory feature of the quantum diffusion approach and we conjecture
that the minimal PDP described in the
following sub-section should be used for simulation of  discrete time
series of experimental events and accompanying them quantum jumps.
An illuminating discussion comparing
the two approaches (i.e. piecewise deterministic vs continuous
diffusion)
 in the domain of quantum optics experiments can be found
in the paper by Wiseman and Milburn$^{(38)}$. A theoretical scheme
which apparently aims at a description of both approaches (but again,
not dealing explicitly with the event space) has
been developed by Barchielli and Belavkin$^{(39)}$.

\subsection{ The Piecewise Deterministic Random Process}
Here we will describe the piecewise deterministic random process (PDP)
on pure states of the total system $Q+C$ that leads to the
Eq.(\ref{eq:rd}) after averaging over paths. It is derived in details in
Ref. $(20)$.

Suppose at $t=0$ the quantum system is described by the state vector
$\psi_0\in\hq$ and the classical system is in the state $\iota_0$. Then
$\psi$ develops according to the equation
\be
\psi(t)={{\exp\left(-iH_{\iota_0}t-{\Lambda\over2}t\right)}\over\Vert
{\exp\left(-iH_{\iota_0}t-{\Lambda\over2}t\right)}\Vert}\psi_0
\ee
while $C$ remains at $i_0$ until jump occurs at some random instant
of time $t_1$.

The time $t_1$ of the jump is governed by the nonhomogeneous Poisson
process that can be described as follows: the probability
$P(t,t+\triangle t)$ for the jump to occur in the time interval
$(t,t+\triangle t)$, provided it did not occurred yet, is given by the
formula
\be
P(t,t+\triangle t)=1-\exp \left(-\int_t^{t+\triangle t}
\lambda(\psi(s))\right)ds ,\label{eq:rt}
\ee
where
\be
\lambda(\psi)=(\psi, \Lambda \psi)\, .
\ee
When the jump occurs at $t=t_1$, then $C$ jumps from $i_0$ to one of the
states $\alpha(i_0)$, say, to $\alpha_1(i_0)$, while quantum state
vector jumps at the same time from its actual value $\psi(t_1)$ to
$\psi_1={F_{\alpha_1} \psi(t_1)}/{\Vert F_{\alpha_1} \psi(t_1)\Vert}$,
and the process starts again.

The probability $p_\alpha$ of choosing a particular value $\alpha$ is
given by
\be p_\alpha={\Vert F_\alpha\psi(t_1)\Vert^2\over\lambda(\psi(t_1))}.
\ee
\medskip\\
\noindent {\bf Remark 5}
It should be noticed that we never assumed that the different
$F_\alpha$-s commute. What can be expected if they do not commute is
instability: instead of stabilizing, the sequence $\alpha_n(i_0)$ covers
part of $\cs$ in a chaotic way.\footnote{Of course the actual behavior
depends essentially also on the relation of the two time-scales: the one
given by the energy spectrum of $\psi$, and the other provided by the
jump rate function.}

\section{Multiparticle Cloud Chamber Model}
As an illustration of the above formalism, in this section we will
consider a specific class of physical models, namely non--relativistic
cloud chamber models.\footnote{The Reader may wish to compare our
simple model with a Hamiltonian theory developed by Belavkin
and Melsheimer in Ref. (39a). The homogeneous Poissonian sampling law is
an assumption qin their paper.}   One such model has been already
described$^{(20)}$, but only for one particle systems. The formalism
developed in the previous section allows us to discuss here a  more
general case.\medskip

We start with a set $E$, thought of to be the physical {\sl space}. We
suppose $E$ is a measurable space, endowed with some measure. To
simplify the notation we will denote the points of $E$ by $x$, an we
will write $dx$ for the corresponding measure. We take ${\cal H}_1$ to
be the Hilbert space of square integrable functions on $E$.\footnote{Or,
more generally, of square integrable sections of a Hermitian vector
bundle over $E$. This will happen for non--scalar particles.} For
instance we can think of $E=R^3$ and ${\cal H}_1=L^2(E,d^3x)$ -- see the
example below. We define $\hq$ to be the Fock space over ${\cal H}_1$,
that is:
\be \hq = \oplus_{n=0}^\infty {\cal H}_n,
\ee
where
\be
{\cal H}_n=\otimes^n {\cal H}_1.
\ee
We denote by $\hq^{\pm}$ the symmetric (Boson) and the antisymmetric
(Fermion) subspace of $\hq$ respectively. We denote by $N(x)$ the number
operator density:
\be
\left(N(x)\psi\right)(x_1,\ldots ,x_n)=\sum_{i=1}^n\delta(x-x_i)
\psi(x_1,\ldots ,x_n)\,
\ee
where $\delta(x-y)$ is the Dirac measure:
$$\int_E f(x)\delta(x-y) dvol(x)=f(y)\, .$$
Then $\hq^{\pm}$ are invariant under $N(x)$.

Let now $E_d$ be another measurable set. The points $a$ of $E_d$ will
parameterize detectors. $E_d$ can be finite (for instance, just one
point), it can be infinite denumerable (e.g. an infinite lattice in
$R^3$) or, if $E=R^3$, $E_d$ can be a lower dimensional submanifold of
$E$ (a string, a surphace). For definitness we can think of $E_d$
as an an open subset of $E$ (in particular, we can take $E_d=E$).
Generalization to other cases presents no problem. One needs, in
general, some measure on $E_d$. The points $a\in E_d$ will play the
role of $\alpha$-s of the previous section.

For each $a\in E_d$, that is for each detector, let there be given a
function $f_a(x)$ on $E$. The physical interpretation is that $f_a(x)$
describes sensitivity of the detector located at the point $a$. For
instance, if $E$ is a Riemannian manifold, we could take for $f_a$ a
Gaussian function of the geodesic distance from $a$. In that case we
would have to provide two parameters (that can depend on $a$) - the
height and the width of the Gaussian function. The height would then be
approximately inversely proportional to the square root of the response
time of the local counter, while the width - to its spatial extension.\\
A point limit corresponds to $f_a^2(x)\mapsto \lambda\delta(x-a)$.

We choose now the \lqq properties\rqq\,  to be given by the
operators\footnote{This choice is equivalent to the one proposed
implicitly by Gisin$^{(40)}$. It is also the same definition as in Ref.
${(36)}$. Notice
howere change in the notation.}
\be F_a=\int_E f_a(x)N(x)dvol(x).
\ee

Our classical system is, as in the Ref. $(20)$ a continuous medium of
2-state detectors which, at each point $a\in E_d$, can be in one of
their two states: \lqq on--state\rqq, represented by ${1\choose0}$, or
\lqq off--state\rqq, represented by ${0\choose1}$. We will consider only
those
configurations of the detector medium which are \lqq on\rqq\,  at a finite
number
of points. Thus the space $\cs$ of pure states of the classical system
is, in our case, the same as the class of all finite subsets of
$E_d$.\footnote{Somewhat more detailed discussion can be found in Ref.
$(20)$}. We will use the letter $\Gamma$ to denote its generic point.
Thus each $\Gamma$ is a finite subset of $E_d$. It represents state of
the detector medium characterized by the fact that the detectors at the
points of $\Gamma$ are \lqq on\rqq\,  while those outside of $\Gamma$ are
\lqq off\rqq.
The symbol $\Gamma$ plays just the role of $\iota$ of the previous
section.

According to our general strategy described before, we consider now the
total system, consisting of the quantum system represented by the Fock
space $\hq$ and of the classical system -- the detectors. Thus states of
the total system are given by families $\{\rho_\Gamma\}_{\Gamma\in\cs}$
such that $\sum_\Gamma \tr (\rho_\Gamma)=1$. The symbol $\sum_\Gamma$
has to be understood in a generalized sense: it is sum over a discrete
index that is the number of points in $\Gamma$, and integral over the
different configurations that these points can take in $E_d$.

What remains to be specified to fix our model, and to apply the results
of the previous section, are the transformations $a:\Gamma\longmapsto
a(\Gamma)$. In our case, when $E_d$ is a subset of $E$, there is a
natural choice: each $a$ flips the detector state at $x=a$. We can write
it also as $a(\Gamma)=\{a\}{\scriptstyle\triangle} \Gamma$, where
${\scriptstyle\triangle}$ denotes the symmetric difference.\footnote{Or,
in words, $a(\Gamma)=
\Gamma\setminus\{a\}$ if $a\in \Gamma$, and $a(\Gamma)=\Gamma\cup\{a\}$
if $a\notin \Gamma$.} It is evident now that $a(a(\Gamma))=\Gamma$, and
so the results of the previous section are applicable. It should be
noticed that the operators $F_a$ commute with the number operator and
with particle permutations. It
follows that if the quantum Hamiltonian preserves the particle number
and particle statistics, then the PD random process preserves them too.
\medskip\\
\noindent{\bf Remark 6}
With real detectors it is realistic to assume that after registering an
event, a certain \lqq dead\rqq\,  interval of time must lapse before they
can
be active again. Moreover, for this time interval the quantum Hamiltonian
may change in the region occupied by the detector. It is very easy to
include such data into our PDP  description. However, the modified
process cease to be Markovian. We can make it Markovian again by
replacing the space of events $\cs$ with the space of \lqq histories\rq\rq.
Then
our framework can be applied again provided we allow that $\cs$ may be
different for each $t$. That is possible and allowing for such a
generalization would not change much our discussion.

\subsection{Example: Spontaneous Localization Model}
We take the simplest case, that of a passive, homogeneous medium$^{(20)}$ in
$E=E_d=R^3$. For the functions $f_a$ we take,
as in GRW$^{(41)}$, the Gaussian functions:
\be f_a(x)=\lambda^{1/2}\left({\alpha\over\pi}\right)^{3\over 2}
exp\left({-\alpha (x-a)^2}\right), \ee where $\alpha$ is the parameter
determining the width of the Gaussian function.\\ Let us consider the
$n$-particle subspace of the Bosonic Fock space. We are working in
position representation, so that pure quantum states are described by
symmetric wave functions $\psi(x_1,\ldots ,x_n)$, with the normalization
\be
 \int \vert \psi(x_1,\ldots ,x_n)\vert^2 dx_1\ldots dx_n =1 .
 \ee
 The operators $F_a$ are easy to compute:
 \be
 \left(F_a\psi\right)(x_1,\ldots ,x_n)=
 \left(f_a(x_1)+\ldots +f_a(x_n)\right)\psi(x_1,\ldots ,x_n) .
 \ee
By simple Gaussian integration we find then the operator
$\Lambda=\int F_a^2 da$. It acts as
\be
\Lambda=\lambda(n+2G),
\ee
where $G$ is the multiplication operator by the function
\be G(x_1,\ldots ,x_n)=\sum_{i<j\le n}\exp\left(-{\alpha\over
4}(x_i-x_j)^2\right).
\ee
The evolution equation (\ref{eq:rqd}) for the effective statistical
state of the quantum system is now exactly the same as in Ghirardi,
Pearle, Rimini paper$^{(41)}$.
Suppose we have free particle dynamics for the quantum particle. Then,
between jumps, the wave function evolves, {\sl up to a normalization
factor} according to the modified Schr\"odinger
equation:\footnote{We omit the scalar dumping term which cancels
out after normalization.}
\be
{{d\psi(x_1,\ldots, x_n,t)}\over
{dt}}=\left({{i\hbar}\over{2m}}\sum_{i=1}^n\triangle_{x_i}
-\lambda G\right)\psi(x_1,\ldots ,x_n,t).
\ee
It is seen that the extra term in the Schr\"odinger equation will dump
the wave function at the coinciding points $x_i\approx x_j$. The degree
of this dumping will depend on the relation of the parameters
$\alpha$,$\lambda$, and on the energy spectrum of the wave function.\\
The rate of jumps is not constant. The probability for a jump to happen in
the infinitesimal time interval $(t,t+dt)$, provided it did not occur
yet, is, according to the formula (\ref{eq:rt}):
\be
P(t,t+dt)=\lambda(t)dt=
\lambda\left(n+(\psi_t,G,\psi_t)\right)dt.
\ee
Here again the coinciding points contribute to a faster reduction
rate.\\ The probability density of triggering a responce from the
detector at some point $a$, given that the event occurs at time $t_1$,
is given by
\be
p_a = \Vert F_a\psi\Vert^2/\lambda(t_1).
\ee
The presence of mixed terms obscures the analysis of most probable
behavior. It will depend on the values of the parameters and on the
initial wave function. Our formulas provide the frame for a numerical
simulation in those ranges where qualitative analysis would provide no
guide.\\

\section{Summary and Conclusions}
We outlined a general philosphy behind the \lqq Event-Enahanced Formalism
of Quantum Theory\rqq \footnote{A recent paper by
Landsman$^{(42)}$ may be helpful for comparing our philosophy
to other approaches, in particular to
the \lqq environment--induced superselection rules\rqq\,  of
Zurek$^{(43)}$.} and we exemplified it with a multiparticle cloud
chamber model. In a particular case what we obtained this way covers
the improved spontaneous localization model. Although we get the same
time evolution equation for the effective statistical state of the quantum
system as Ghirardi, Pearle and Rimini, there are differences too. First of
all we see that the time evolution of the quantum state separates
only in a particular, simple case. Second, for the random process we obtain
the minimal PDP that takes place on pure states of $Q+C$. That process, for
the cloud chamber model, is easy to verify experimentally and to simulate
numerically (apart of the computer power which is necessary for  numerical
solutions of multiparticle,
multidimensional, time-dependent and non-unitary Schr\"odinger equation).
In a particular case, when the quantum Hamiltonian does not depend on
the state of the classical detectors, the quantum evolution separates
and our process coincides with the Monte Carlo Wave Function algorithm
as described in Refs. (31-34).\\
One important  idea of the event-enhanced formalism, namely that the enhanced
quantum theory provides its own interpretation, was not discussed in this
paper.  It will be described in the forthcoming paper by Ph. Blanchard and
the present author$^{(44)}$.
\vskip10pt
\noindent
{\bf Acknowledgements}\\
I would like to thank the Erwin Schr\"odinger Institute in Wien
for hospitality and support during the initial work on this paper.
Thanks are
due to H. Narnhofer for reading the manuscript. Thanks are also due to
A. Amann, F. Benatti and N. Gisin for
useful comments that shaped the final form of this paper.
I owe lot of thanks to  H.P. Stapp for correspondence
and for his criticism concerning arbitrariness of $Q+C$ seperation. I tried
to answer at least some of his objections.
I also thank to R. Haag for a discussion that explained to me
better his point on view on the world event--process.
The final writup was done at BiBoS, Bielefeld. Without
discussions with Ph. Blanchard this paper would never be born.
\vskip10pt
\noindent
{\large{\bf References}}

\begin{description}
\item{1.} Schr\"odinger, E. : \lqq The Philosophy of Experiment\rqq,
{\sl Il Nuovo Cimento, }{\bf 1} (1955), 5-15
\item{2.} Stapp, H. P. : \lqq Bell's Theorem and World Process\rqq,
{\sl Nuovo Cimento} {\bf 29} (1975) 270--276
\item{3.} Stapp, H. P. : \lqq Theory of Reality\rqq, {\sl Found. Phys.}
{\bf 7} (1977) 313--323
\item{4.}Stapp, H. P. : \lqq Mind, Matter and Quantum
Mechanics\rqq, Springer Verlag, Berlin 1993
\item{5.} Stapp, H. P. : \lqq Solution of the Infrared Problem\rqq,
{\sl Phys. Rev. Lett.} {\bf 50} (1983) 467--469
\item{6.}  Chew, G. F. : \lqq Gentle Quantum Events and the Source
of Explicate Order\rqq, {\sl Zygon} {\bf 20} (1985) 159--164
\item{7.} Shimony, A. : \lqq Events and Processes in the Quantum World\rqq,
in
{\sl Quantum Concepts in Space and Time}, Ed. R. Penrose and C. J.
Isham, Clanderon Press, Oxford 1986
\item{8.} Haag, R. : \lqq Events, histories, irreversibility\rqq, in
{\sl Quantum Control and Measurement}, Proc. ISQM Satellite Workshop,
ARL Hitachi, August 28--29, 1992, Eds. H. Ezawa and Y. Murayama,
North Holland, Amsterdam 1985
\item{9.} Haag, R. : \lqq Fundamental Irreversibility and the Concept of
Events\rqq, {\sl Commun. Math. Phys.}{\bf 132} (1990) 245--251
\item{10.}  Barut, A.O. : \lqq Quantum Theory of Single Events: Localized
De Broglie Wavelets, Schr\"odinger Waves, and Classical Trajectories\rqq,
{\sl Found. Phys.} {\bf 20} (1990), 1233--1240
\item{11.}  Bell,  J. : \lqq Against measurement\rqq, in
{\sl Sixty-Two Years of Uncertainty. Historical, Philosophical and
Physical Inquiries into the Foundations of Quantum Mechanics}, Proceedings
of a NATO Advanced Study Institute, August 5-15, Erice, Ed. Arthur I. Miller,
NATO ASI Series B vol. 226 , Plenum Press, New York 1990
\item{11a.} Bell,  J. : \lqq Towards an exact quantum mechanics\rqq,  in
{\sl Themes in Contemporary Physics II.  Essays in honor of Julian
Schwinger's 70th birthday},  Deser,  S. ,  and Finkelstein,  R. J.  Ed. ,
World Scientific,  Singapore 1989
\item{12.} Rauch, H. : \lqq Superposition Experiments in Neutron
Interferometry\rqq, Preprint, Atominstitut \"Osterreichischen
Universit\"aten, Wien 1994
\item{13.} Zawisky, M., H. Rauch and Y. Hasegawa, \lqq Contrast Enhancement
by Time -- Selection in Neutron -- Interferometry\rqq, Preprint,
Atominstitut \"Osterreichischen
Universit\"atenWien 1994, to
appear in {\sl Phys. Rev. A}
\item{14.} Blanchard,  Ph.  and Jadczyk,  A. : \lqq On the interaction
between classical and quantum systems\rqq, {\sl Phys. Lett. }{\bf A
175} (1993), 157--164
\item{15.}  Blanchard,  Ph.  and Jadczyk,  A. : \lqq Classical and quantum
intertwine\rqq,  in {\sl Proceedings of the Symposium on Foundations of
Modern Physics},  Cologne,  June 1993,  Ed.  P.  Mittelstaedt,  World
Scientific  (1994),  hep--th 9309112
\item{16.} Blanchard,  Ph.  and Jadczyk,  A. : \lqq Strongly coupled
quantum and classical systems and Zeno's effect\rqq, {\sl
Phys. Lett. }{\bf A 183} (1993), 272--276
\item{17.}  Blanchard,  Ph.  and Jadczyk,  A. : \lqq From quantum
probabilities to classical facts\rqq,  in {\sl Advances in Dynamical
Systems and Quantum Physics},  Capri,  May 1993,  Ed.  R.  Figario,  World
Scientific  (1994), hep--th 9311090
\item{18.} Blanchard,  Ph.  and Jadczyk,  A. :  \lqq How and
When Quantum Phenomena Become Real\rqq,  to appear in Proc.
Third Max Born Symp.  \lqq Stochasticity and Quantum Chaos\rqq,
Sobotka,  Eds.  Z.  Haba et all. ,  Kluwer Publ.
\item{19.}  Jadczyk, A. : \lqq Topics in Quantum Dynamics\rqq, Preprint
CPT--Marseille 94/P.3022, also BiBoS 635/5/94, hep--th 9406204
\item{20.}  Jadczyk, A. : \lqq Particle Tracks, Events and Quantum
Theory\rqq,  preprint RIMS, hep-th 9407157
\item{21.} Blanchard, Ph. and Jadczyk, A. : \lqq Event-Enhanced Formalism of
Quantum Theory or Columbus Solution to the Quantum Mesurement Problem\rqq,
to appear in Proc. Int. Workshop on  Quantum Communication
and Measurement, Nottingham, 11--16 July 1994
\item{22.} Stapp, H. P. :\lqq The Integration of Mind into Physics\rqq,
talk at
the conference {\sl Fundamental Problems in Quantum Theory.} Univ. of
Maryland at Baltimore, June 18-22, 1994. Auspices of the NYAc.Sci.
Honoring Professor John Archibald Wheeler.
\item{23.} Heisenberg, W. : {\sl The Physical Principles of Quantum
Mechanics}, The University of Chicago Press, Chicago 1930
\item{24.} Spiller, T.P., Clark, T.D., Prance, R.J., and Prance, H. :
\lqq The adiabatic monitoring of quantum objects\rqq, {\sl Phys. Lett.}
{\bf A 170} (1992), 273-279
\item{25.} Gorini, V. , Kossakowski, A. and Sudarshan, E. C. G. :
\lqq Completely positive dynamical semigroups of N--level systems\rqq,
{\sl J. Math. Phys. }{\bf 17} (1976), 821--825
\item{26.} Lindblad,  G. : \lqq On the Generators of Quantum Mechanical
Semigroups\rqq, {\sl Comm. Math. Phys. }{\bf 48} (1976), 119--130
\item{27.} Christensen,  E.  and Evans,  D. : \lqq Cohomology of
operator algebras and quantum dynamical semigroups\rqq,  {\sl J.  London.
Math.  Soc. }  {\bf 20} (1978), 358--368
\item{28.} Fredkin, E., and Toffoli, T. : \lqq Conservative Logic\rqq,
{\sl Int. J. Theor. Phys.} {\bf 21} (1982), 219--254
\item{29.} Davis, M. H. A. : {\sl Lectures on Stochastic Control and
Nonlinear Filtering}, Tata Institute of Fundamental Research, Springer
Verlag, Berlin 1984
\item{30.} Davis, M. H. A. : {\sl Markov models and optimization},
Monographs on Statistics and Applied Probability, Chapman and Hall,
London 1993
\item{31.} Carmichael, H. : {\sl An open systems approach to quantum
optics}, Lecture Notes in Physics m 18, Springer Verlag, Berlin 1993
\item{32.} Dalibard, J. , Castin, Y. and M{\o}lmer K. : \lqq Wave--function
approach to dissipative processes in quantum optics\rqq, {\sl Phys. Rev.
Lett. }{\bf 68} (1992), 580--583
\item{33.} Dum, R. , Zoller, P. , and Ritsch, H. : \lqq Monte Carlo
simulation of the atomic master equation for spontaneous emission\rqq, {\sl
Phys. Rev. }{\bf A45} (1992), 4879--4887
\item{34.} Gardiner, C. W. , Parkins, A. S. , and Zoller, P. :
\lqq Wave--function stochastic differential equations and quantum--jump
simulation methods\rqq, {\sl Phys. Rev. } {\bf A46} (1992), 4363--4381
\item{35.} Pearle, P. : \lqq Combining stochastic dynamical state-vector
reduction with spontaneous localization\rqq, {\sl Phys. Rev.}{\bf A 39}
(1989), 2277-2289
\item{36.} Ghirardi, G. C., Pearle, P., Rimini, A. : \lqq  Markov processes
in Hilbert space and continuous spontaneous localization of systems of
identical particles\rqq, {\sl Phys. Rev.} {\bf A 42} (1990), 78-95
\item{37.} Gisin, N., and Percival, I. C. : \lqq The quantum state
diffusion picture of physical processes\rqq, {\sl J. Phys.} {A 26}
(1993), 2245--2260
\item{38.} Wiseman, H.M., and Milburn, G.J. : \lqq Interpretation of
quantum jump and diffusion processes illustrated on the Bloch
sphere\rqq, {\sl Phys. Rev. }{\bf A 47} (1993), 1652--1666
\item{39.} Barchielli, A. and Belavkin, V.P. : \lqq Measurement
continuous in time and {\sl a posteriori} states in quantum
mechanics\rqq, {\sl J. Phys.} {\bf A 24} (1991), 1495--1514
\item{39a} Belavkin, V. and Melsheimer, O. : \lq A Hamiltonian
theory for continuous reduction and spontaneous localization\rq,
Preprint Centro Vito Volterra N. 139, May 1993
\item{40.} Gisin, N. : \lqq Stochastic quantum dynamics and relativity\rqq,
{\sl Helv. Phys. Acta} {\bf 42} (1989), 363--371
\item{41.} Ghirardi, G.C., Rimini, A. and Weber, T. : \lqq An Attempt at a
Unified Description of Microscopic and Macroscopic Systems\rqq, in {\sl
Fundamental Aspects of Quantum Theory}, Proc. NATO Adv. Res. Workshop,
Como, Sept. 2--7, 1985, Eds. Vittorio Gorini and Alberto Frigerio, NATO
ASI Series B 144, Plenum Press, New York 1986, pp. 57--64
\item{42.} Landsman, N.P. : \lqq Observation and superselection in quantum
mechanics\rqq, to appear in {\sl Studies in History and Philosophy of
Modern Physics} (1995), Preprint DESY 94-141, August 1994
\item{43.} Zurek, W.H. : \lqq Environment--induced superselection rules\rqq,
{\sl Phys. Rev.} {\bf D26} (1982), 1862--1880
\item{44.} Blanchard, Ph. and Jadczyk, A. : \lqq Event-Enhanced Quantum
Theory\rqq, in preparation
\end{description}

\end{document}